\documentclass[final,5p,times]{elsarticle}

\makeatletter
\def\ps@pprintTitle{%
 \let\@oddhead\@empty
 \let\@evenhead\@empty
 \def\@oddfoot{}%
 \let\@evenfoot\@oddfoot}
\makeatother

\usepackage{amssymb}

\usepackage{url}
\usepackage{breakurl}
\usepackage[breaklinks]{hyperref}

\begin{document}

\def\ep{\mathbf{p}}
\def\eq{\mathbf{q}}
\def\eu{\mathbf{u}}
\def\ev{\mathbf{v}}
\def\ex{\mathbf{x}}
\def\ey{\mathbf{y}}
\def\ec{\mathbf{c}}

\begin{frontmatter}



\title{Adaptive Mesh Booleans}
      

\author[add1]{Ryan Schmidt}
\ead{ryan.schmidt@autodesk.com}
\author[add1]{Tyson Brochu}
\ead{tyson.brochu@autodesk.com}

\address[add1]{Autodesk Research}

\begin{abstract}

We present a new method for performing Boolean operations on volumes represented as triangle meshes. In contrast to existing 
methods which treat meshes as 3D polyhedra and try to partition the faces at their exact intersection curves, we treat meshes as 
adaptive surfaces which can be arbitrarily refined. Rather than depending on computing precise face intersections, our approach
refines the input meshes in the intersection regions, then discards intersecting triangles and fills the resulting holes
with high-quality triangles. The original intersection curves are approximated to a user-definable precision,
and our method can identify and preserve creases and sharp features. Advantages of our approach include the
ability to trade speed for accuracy, support for open meshes, and the ability to incorporate tolerances to handle
cases where large numbers of faces are slightly inter-penetrating or near-coincident.
\end{abstract}


\end{frontmatter}


\begin{figure*}[htb]
\centering
\includegraphics[width=.99\linewidth]{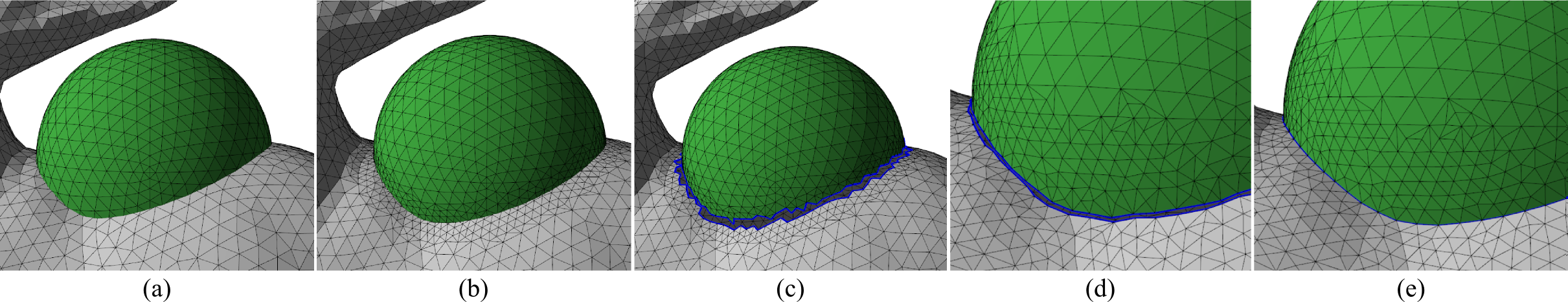}
\caption{\label{teaser} A Boolean union operation performed using our method. Two intersecting meshes (a) are given as input. 
   The meshes are (b) refined in the intersection region, then (c) triangles in the neighborhod of the intersection are discarded.
   The resulting hole is filled (d,e) using an adaptive front-marching strategy that precisely merges the two open boundary loops.  }
\label{fig:teaser}
\end{figure*}

\section{Introduction}

Boolean composition is one of the most basic concepts in geometric modeling.
The notion of union, intersection, or difference of two volumes can be intuitively understood.
This conceptual simplicity is in stark contrast to the complexity of implementations of these operators, which remain weak links in modern CAD tools.
The challenge is that most CAD tools in fact rely on surface or boundary representations (B-reps), such as NURBS patches or triangle meshes.
As a result, volume composition often requires the computation of intersections between thousands, or millions, of surface patches.
Geometric intersection is numerically challenging, and a single wrong predicate result can render the operation a failure.

Our focus is Boolean operators between triangle meshes. 
Existing works treat meshes as 3D polyhedra, and attempt to find intersection curves that cut the input faces.
Our premise, on the other hand, is that when using meshes as high-resolution representations 
of smooth surfaces, it is not necessary to precisely cut each face. 
Instead we propose that, given sufficient mesh density, no individual face is
particularly important, and we are free to locally re-mesh the surface in the course of any operation. 
We will refer to this special case of triangle mesh as an \emph{adaptive mesh surface}.

Adaptive mesh surfaces have been used in various modeling contexts.
For example, remeshing surfaces during simulation is common in the physically-based animation of 
cloth, deformable solids, and fluids~\cite{Narain12}.
Adaptive remeshing is also used in fair surface design~\cite{Schneider01} and deformation~\cite{Dunyach13}.
Recently, interactive design tools have utilized adaptive meshes to provide
more intuitive interactions, such as push/pull deformation~\cite{Stanculescu13} and 
3D sculpting~\cite{Stanculescu11}. Autodesk Meshmixer~\cite{Meshmixer15}
uses similar adaptive mesh strategies in several of its modeling tools.

We present a novel approach to Booleans suitable for use with adaptive mesh surfaces.
Rather than cutting triangles, our method simply adds more, deletes enough to separate
the input surfaces into disjoint patches, and then stitches the patches back together. 
Our stitching algorithm also takes advantage of mesh adaptivity.
Although this strategy only produces approximate Booleans, by constraining the
free boundaries we can find intersection curves that are accurate up to a pre-defined tolerance, or trade time for precision.
This process is illustrated in Figure~\ref{teaser}. 
We can also preserve crease edges in the input surfaces via constraints,
and our method can be adapted to produce approximate Booleans that can be applied to 
meshes of the same shape but varying triangulation.


To summarize, contributions of this paper include:
\begin{enumerate}
\item A method for connecting pairs of open mesh boundaries with high-quality triangles
\item A technique to preserve sharp features during mesh refinement and hole filling
\item A new approach to mesh Boolean operations based on mesh refinement and constrained hole filling
\end{enumerate}

\section{Related Work}

In many computer graphics contexts, we represent solids via their boundaries (often called B-Reps).
Computing Boolean operations on B-Reps has long been a focus of CAGD research.  
For surfaces represented via parametric patches (widely used in commercial CAD systems) robust 
implementations are available in commercial solid-modeling kernels such as ACIS~\cite{ACIS}.
However, these approaches are not designed for high-resolution polygonal meshes, and perform extremely poorly in such cases.

The CGAL library supports robust Boolean operations on Nef polyhedra~\cite{Hachenberger15} 
with exact geometric computation~\cite{Granados03}. 
This precision comes with a cost, with Hachenberger et al.~\cite{Hachenberger05} quoting runtimes of 
hundreds of seconds for relatively simple models.
Arbitrary closed meshes can be converted to Nef polyhedra~\cite{Nef78}, however this can
significantly change the tessellation of the input surface in planar regions outside the intersection region.

BSP-based methods are highly effective for mesh Booleans. 
With careful design of predicates, provably-robust methods have been presented~\cite{Bernstein09}.
Campen and Kobbelt~\cite{Campen10} extended this technique, improving
performance with an adaptive octree and fixed-precision arithmetic.
Wang and Manocha~\cite{Wang13} present a fast and robust technique
for extracting an output mesh from a BSP-tree.
However, the output mesh is again completely re-tessellated.
This is problematic in many contexts where the input meshes may have properties
bound to geometric elements, such as UV-maps. 

Another class of strategies involves using each input mesh to cut the faces of the other. 
The partitioned objects are then stitched along the new boundary loops and mesh components are discarded
according to the type of Boolean operation being performed. 
Publicly-available libraries Cork~\cite{Cork} and Carve~\cite{Carve} take this approach.
Xu and Keyser~\cite{Xu13} propose one such approach, and recently
Barki et al.~\cite{Barki15} presented a method that extends this approach to
non-manifold input, by handling degenerate configurations in a systematic
way (usually the downfall of this approach). Their method is both highly
robust and performant, but does require closed meshes.

With shape representations such as voxels, level sets, or Layered Depth Images~\cite{Zhao11},
robust Boolean operations are much simpler to compute.
However these techniques require discretization of input meshes, which can cause the loss of sharp features and small details. 
Hybrid approaches have been developed which limit discretization (and hence resampling) 
to the neighborhods around intersection contours~\cite{Wang11,Pavic11}.
Although highly robust, these methods have a dependence on the volumetric discretization resolution.
BSP-trees can be used to create precise implicit representations of arbitrary polyhedra~\cite{Fryazinov11},
which can be trivially composed using functional operators.
However this is very expensive for high-resolution meshes, and output mesh
extraction again involves global resampling and potentially the loss of sharp features.

Various other mesh processing techniques have been developed to provide ``Boolean-like'' behavior.
For example Bernstein and Wojtan~\cite{Bernstein13} present a method for adaptively merging
meshes as they collide. Chentanez et al.~\cite{Chentanez15} approximate Boolean union when intersections are
detected during mesh-based fluid surface tracking. 
Similar to our approach, their method deletes overlaps and fill the gaps.
However rather than a simple polygon fill, our method uses adaptive front marching to closely approximate
the intersection curves and can preserve sharp features on the input.

\section{Method}
\label{secMethod}

In this section we present our new approach to adaptive mesh Booleans.
The general strategy is to build a mesh ``zippering'' algorithm on top of 
a mesh refinement which is used to create an approximate Boolean
operator. Constraints are added to the refinement and zippering to increase accuracy.

\subsection{Mesh Refinement}
\label{secRefinement}

Local mesh refinement (or remeshing) is a fundamental operation in an adaptive mesh surface.
Refinement is used not only to adapt the current mesh to higher or lower sampling densities, but
also to optimize the shape and distribution of triangles for a given computation. 
Many approaches have been presented in the literature, in this section we describe our method.

\begin{figure}[htb]
\centering
\includegraphics[width=\linewidth]{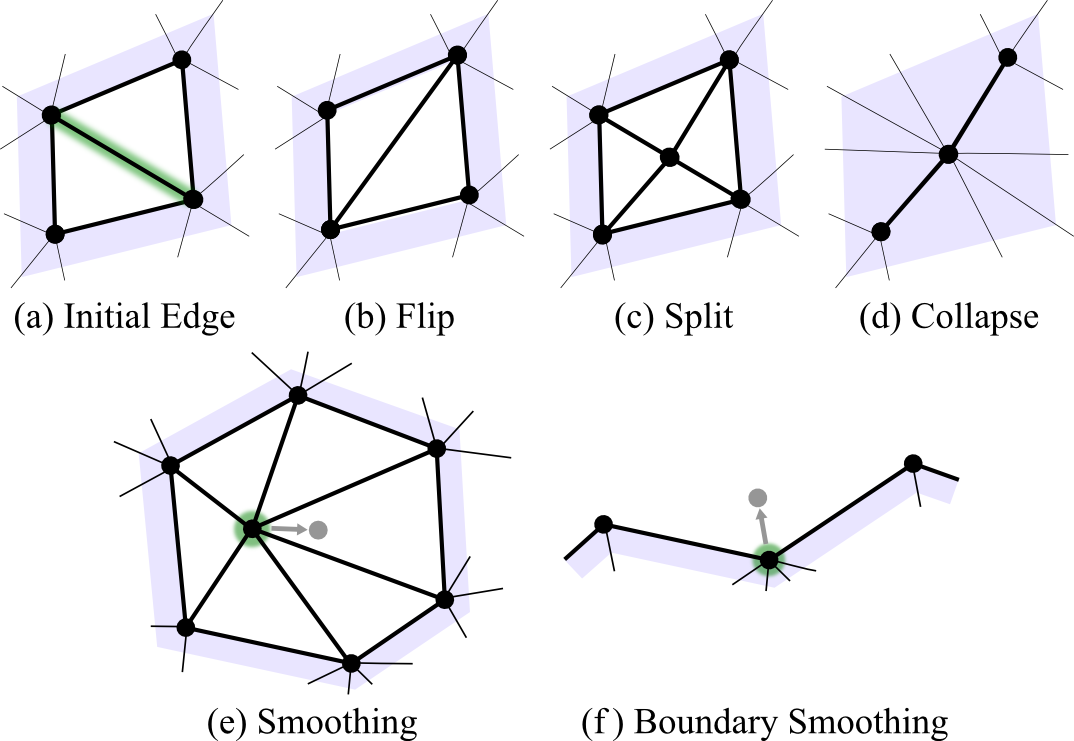}
\caption{\label{figRefinement} Mesh refinement operators}
\end{figure}

We largely follow \cite{Botsch04}, where the mesh is locally modified using well-known edge
split/flip/collapse operators~\cite{Hoppe93}, as well as local vertex-Laplacian smoothing.
These operations are illustrated in Figure~\ref{figRefinement}. 
We apply these operations in sequential passes over the mesh, in the
order Split, Collapse, Flip, Smooth.  
Note that the ordering of operations can have significant impact on the result.
For example, in our current implementation, swapping the Collapse and Flip
steps results in a 25\% reduction in performance.

\subsubsection{Constraints}
\label{sec:constraints}

\begin{figure}[htb]
\centering
\includegraphics[width=\linewidth]{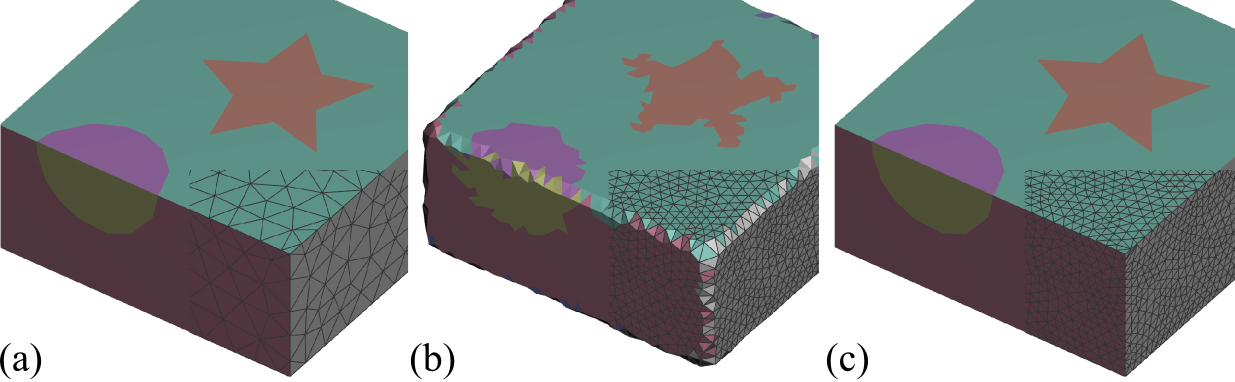}
\caption{\label{figConsRef} The (a) borders of mesh feature regions (here identified by triangle colors) 
will (b) be lost during resampling, unless the (c) edge and smoothing operators are explicitly constrained. }
\end{figure}

Previously we stated that in an adaptive mesh surface, ``no triangle matters''.
However, this is not true of edges. 
Even in a high-resolution mesh, some edges necessarily define feature boundaries. 
In mesh refinement these boundaries must explicitly be preserved, 
otherwise they are certain to be lost, as demonstrated in Figure~\ref{figConsRef}.
Similarly, at open mesh boundaries we must apply various constraints depending
on the desired behaviour. For example, we may wish to exactly preserve boundary
edges, or perhaps preserve the boundary segment but allow resampling.
It is useful to think of the network of feature and border edges as a graph. 
When enforcing constraints, the path between graph nodes may be mutable,
but we must prevent operations that would change the topology of this graph.

\subsubsection{Edge Split}
\label{sec:split}
Assume we have chosen a constant maximum edge length $l_{\max}$. 
In a Split pass, we iterate over the mesh and split any edge longer than $l_{\max}$ by introducing a new vertex. 
The new vertex can be placed at the edge midpoint (linear subdivision) or placed on a curve
estimated using local B{\'e}zier interpolation~\cite{Boschiroli11}.
On feature and boundary edges, only linear subdivision is used.

\subsubsection{Edge Collapse}
\label{sec:collapse}
We define the minimum edge length $l_{\min}$ as a fraction of the maximum length $l_{\max}$.
In the context of our Booleans we use $l_{\min} = 0.4 l_{\max}$. 
(We use different minimum and maximum values for other mesh refinement operations in our larger mesh modeling package.)
In Collapse passes, we iterate over the mesh edges, and if an edge is shorther than $l_{\min}$,
or if the opening angle at either opposing vertex is less than ${\pi}/12$, we attempt to
collapse the edge by replacing it with a single vertex. 
This new vertex is placed at the edge midpoint unless one endpoint of the edge lies on a mesh or feature
border, in which case we collapse to that vertex's position. 
If both vertices of an edge lie on feature constraints, but the edge itself is not a feature
edge, then we cannot collapse this edge as it would change the feature topology.

\subsubsection{Edge Flip}
\label{sec:flip}
Edge flips can be used to normalize vertex valences~\cite{Botsch04}, leading to more regular triangles.
However, we found that when adding feature constraints, this policy can lead to poor-quality
triangles around constraint boundaries.
Instead we use a ``flip-to-shorter'' policy.  
For a given edge, we can form a second edge by connecting the opposing vertices in the two connected triangles.
If this second edge is shorter than the current edge, and the flip does not result in significantly worse aspect
ratios or inverted faces, we perform the flip.
Boundary and feature edges are not flipped.

\subsubsection{Smoothing}
\label{sec:smoothing}

Laplacian mesh smoothing (moving vertices toward the centroid of their neighborhood) is often used to 
improve mesh regularity by driving mesh edges to be equal in length. We use an inverse-area-weighted 
smoothing approach. For each vertex $i$, we first compute the centroid of its neighborhood, $\ec_i$. The 
vertex position is then set to be:
$$
(1 - \alpha A) \ev_i + \alpha A \ec_i
$$
where $A$ is the reciprocal of the mixed area of the vertex \cite{Meyer02}, and $\alpha$ is a user-controlled 
smoothing weight. 

\subsection{Adaptive mesh zippering}
\label{secZipper}

\begin{figure}[htb]
 \centering
  \includegraphics[width=.9\linewidth]{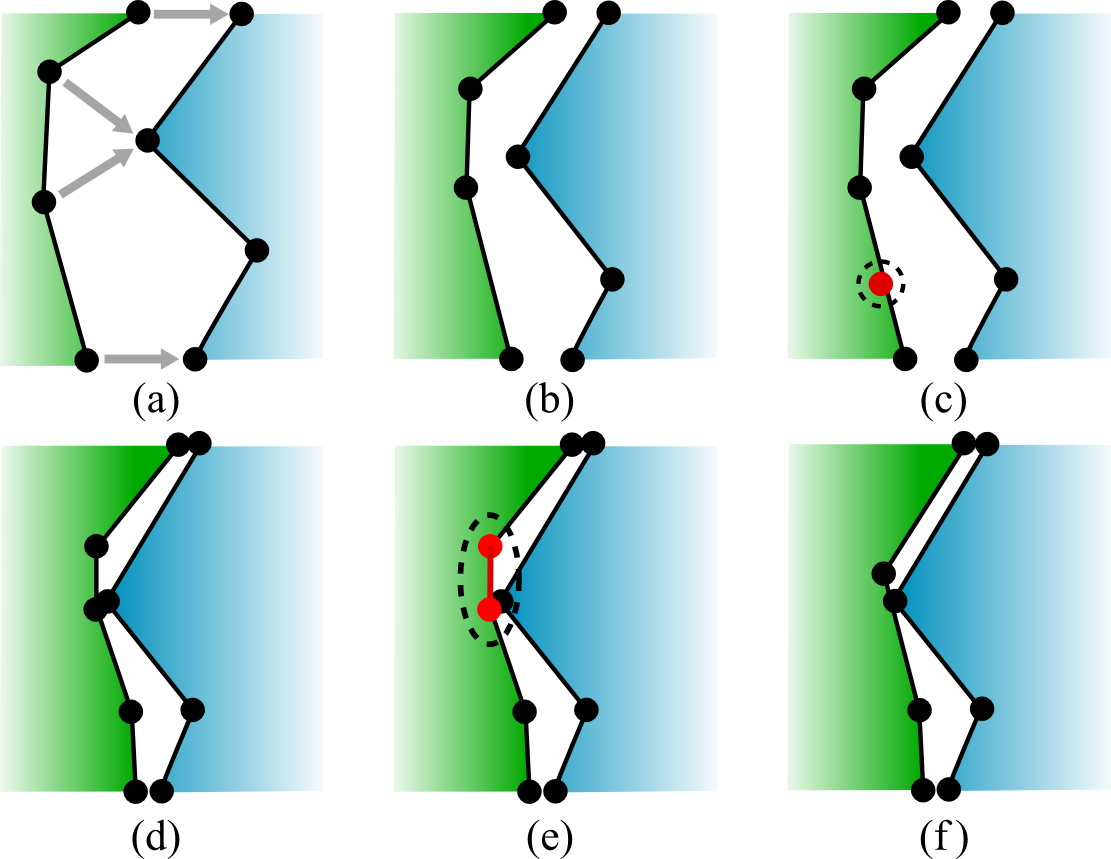}
  \caption{\label{figVVICP} Our adaptive mesh zippering algorithm. 
In the first iteration (a,b) each vertex steps towards its nearest neighbor on the other boundary.  
One edge lengthens (c) to the point where an edge split occurs. 
After the second iteration (d), an edge has become short enough to collapse (c).
At this point Euclidean distance defines a bijective correspondence between the sets of vertices. }
\end{figure}

Our mesh zippering approach is illustrated in figure \ref{figVVICP}. 
Assume we have boundary loops $l_1 = \{\ev_0,\ev_1,\ldots,\ev_n\}$ and $l_2 = \{\eu_0,\eu_1,\ldots,\eu_m\}$.
We can define a function $nearest(l,\ex)$ which returns the vertex in $l$ which is nearest to $\ex$ under Euclidean distance.
To merge the loops, we simultaneously evolve $l_1$ and $l_2$ until $m = n$ and for 
any $i$, $nearest(l_1, nearest(l_2,\ev_i)) = \ev_i$. 
At this point we can construct a trivial bijective correspondence, and the two boundary loops can be merged 
simply by replacing each $\ev_i$ and $nearest(l_2, \ev_i)$ with a new vertex.
The resulting mesh will be manifold in the neighborhood of this boundary.

Our boundary evolution is a basic \emph{iterative closest point} strategy:

\begin{enumerate}
\item For each $\ev_i$ in $l_1$:
\begin{enumerate}
\item Find $\ev^n_i \coloneqq nearest(l_2,\ev_i)$
\item Find the new point $\ev_i^m \coloneqq (1-t) \ev_i + t \ev^n_i$, where $t \leq 0.5$
\end{enumerate}
\item Repeat for each $\eu_j$ in $l_2$
\item Update the positions in $l_1$ and $l_2$ to the new points $\ev_i^m$ and $\eu_j^m$
\item Refine the meshes in the neighborhood of $l_1$ and $l_2$.
\end{enumerate}

Steps 1-3 clearly will converge, but will not produce a bijective one-to-one mapping in the general case.
Multiple $\ev_i$ will likely collapse to a single $\eu_j$, creating degenerate edges in the loop $l_1$.
However, by including a refinement step, any near-degenerate edges will collapse until a single vertex remains.
Similarly, if two vertices are pulled sufficiently far apart, an edge split will introduce a new vertex
between them, dealing with the T-junction case. Figure~\ref{figVVICP} illustrates this process.

\begin{figure}[htb]
 \centering
  \includegraphics[width=\linewidth]{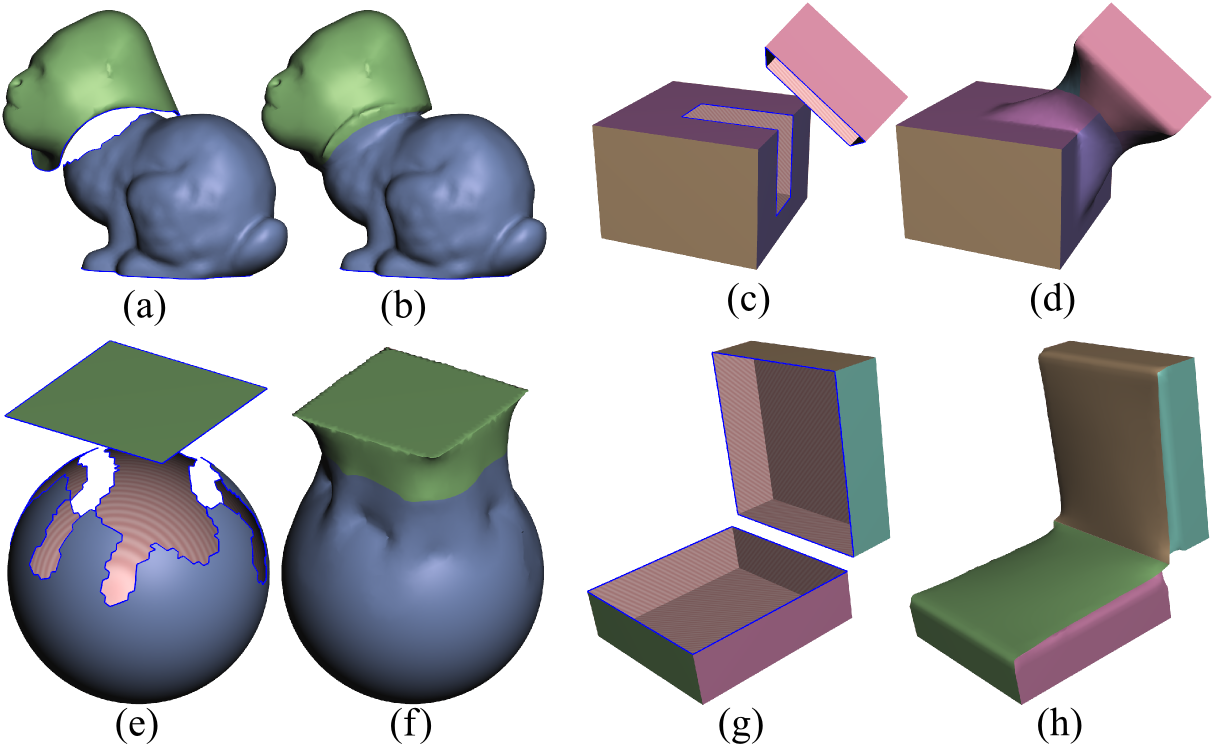}
  \caption{\label{figZipper} Complex examples handled by our adaptive mesh zippering algorithm.
The cases (a,c,e,g) each have two open boundary loops, which are joined in the results (b,d,f,h).
Note that due to our use of basic Euclidean distance, the join border in (h) is non-manifold.
 }
\end{figure}

Although very simple, this strategy is remarkably robust, and can automatically create transitions
between meshes with highly variable resolution. 
Obviously if $l_1$ and $l_2$ have very different shapes and arbitrary spatial orientation, 
the algorithm above cannot guarantee  that the resulting zippered mesh does not have self-intersections.
The loop correspondence produced may be non-manifold in such cases (Figure~\ref{figZipper}).
It is also possible to construct pathological cases where the refinement will not resolve local duplicate triangles. 
However, within the context of our Mesh Booleans, the two loops lie within a bounded distance
that is on the order of the triangle edge lengths, which is ideal. 
This algorithm has been used in the commercial software Autodesk Meshmixer~\cite{Meshmixer15}
for over 3 years and has been highly reliable in practice.

\subsection{Boolean Operation}
\label{secBasicBool}

We assemble our Boolean operation using the techniques described above.
We begin with the simplified case where there is a single intersection between
the two objects, and then explain how our method generalizes to multiple intersections.

Assume we have closed convex input meshes $M_1$ and $M_2$, which have one intersecting region.
The first step is to locate the sets of intersecting triangles. 
A spatial data structure greatly accelerates this search; we use triangles sorted into a sparse octree, 
however we do not claim that this is the optimal acceleration structure.
Note that we do not need to cut the triangles, simply determine if they intersect.
The result is two sets $t_1 \in M_1$ and $t_2 \in M_2$, each of which is topologically
equivalent to either an annulus or a disk. 

The next step is to delete the sets $t_1$ and $t_2$ from $M_1$ and $M_2$, respectively.
If the newly-created open boundaries contain any ``bowtie'' vertices (connected to more than 
one group of adjacent triangles), we remove those vertices and their one-rings, and repeat this step 
until the boundaries are manifold curves.
This produces four separate mesh patches which are disc-shaped regions with open boundaries.
We now need to discard some of these patches. Which to discard depends on the particular
Boolean operation. For a union operation, we would discard the patch of $M_2$ which is
contained \emph{inside} the original $M_1$, and vice-versa. For intersection, we discard the opposite
set of patches, but must also reverse the orientation of the remaining patches. 

In our example case, we are now left with two patches, topologically equivalent
to discs, with nearby open boundaries. The zippering algorithm from the
previous section can be directly applied to merge these open boundaries.
The result is an approximate Boolean composition. 
Figure~\ref{teaser} illustrates this process in action.

Now let us consider cases where there are multiple intersections, and/or non-genus-zero objects.
In both cases, the only added complication is that we will produce more patches and boundary loops,
and we must sort out which loops to zipper together. 
We found that a simple voting scheme suffices, where each loop vertex ``votes'' for its
nearest loop on the other mesh. Loops that agree are paired. 

Similarly, the method does not require closed input meshes. 
We can determine containment statistically: for a given patch, we
cast $N$ random rays from the surface and if more than $N/2$
rays hit the interior side of the containment mesh, this patch is classified as inside.
More advanced techniques such as the generalized winding-number~\cite{Jacobson13}
would further improve this aspect.

\subsection{Robust Boolean}

\begin{figure}[htb]
\centering
\includegraphics[width=\linewidth]{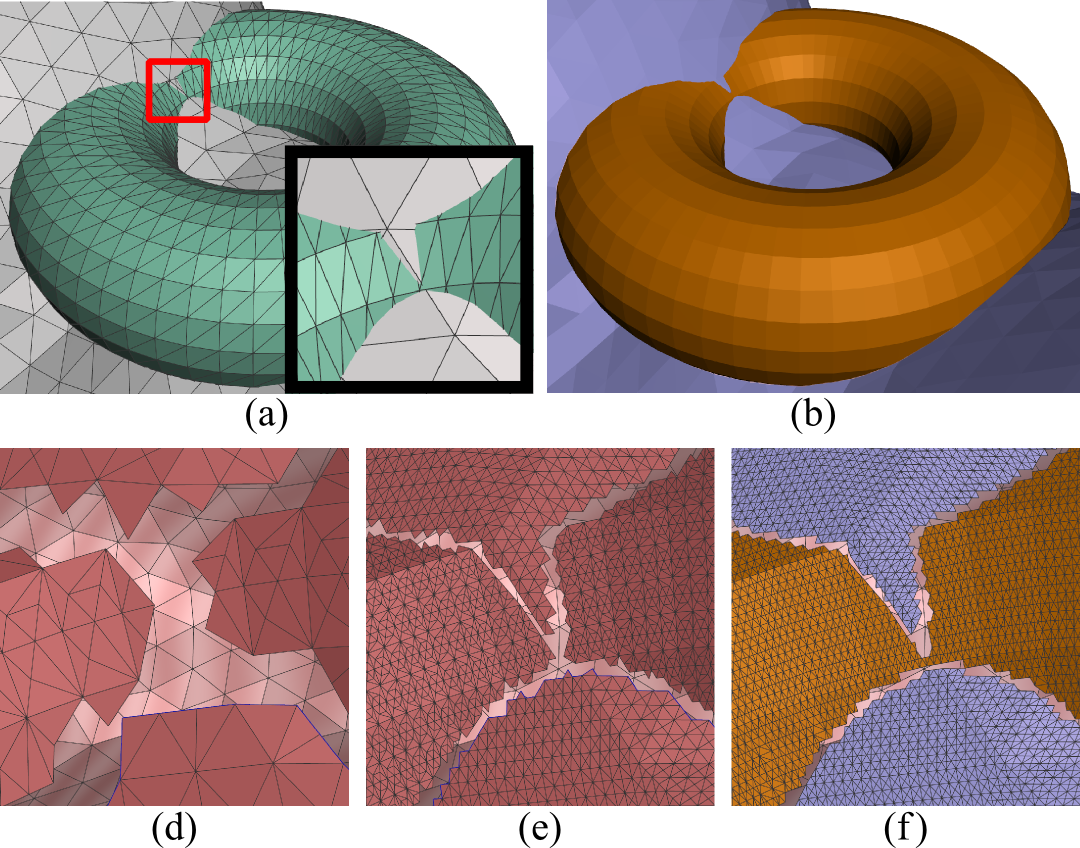}
\caption{\label{figIterations} The torus in (a) is near-coincident with
the grey surface, and requires 8 levels of refinement to produce (b) a successful result.
Bottom row shows levels 2 (c), 7 (d), and 8(e). }
\end{figure} 

In the development above, we noted that the initial intersection regions $t_1$ and $t_2$
needed to be topologically equivalent to annuli. 
Otherwise, we may have different numbers of open boundary loops for $M_1$ and $M_2$ 
and the result will either contain holes or floating patches. 
This is the primary failure mode for our technique. 
It occurs when either the inexact triangle/triangle intersection tests produce inconsistent results due 
to numerical issues, or when the ``feature size''
of the intersection curve is smaller than the current mesh resolution (i.e., there is local undersampling
in the context of the operation we are trying to perform).

Although heuristics and repairs could be applied to handle some cases, a more reliable strategy 
is to apply our adaptivity assumption, and simply repeat the process with increased mesh density.
After finding the initial intersection regions $t_1$ and $t_2$, we grow each region to include 
its one-ring and apply mesh refinement as described above. 
This process may need to be repeated several times, but at some sufficient mesh resolution,
the intersection regions will have the correct topology and the operation will succeed.
This process is illustrated in Figure~\ref{figIterations}.

In practice, we cannot subdivide infinitely, due to both memory and floating-point limitations. 
In our implementation we use a fixed number of iterations (5) and allow the user to add more if necessary.
Note that the previous intersection and refinement steps can be re-used during the next round.

Currently we apply uniform refinement in both the initial intersection regions 
and during the zippering operation. A useful extension would be to adapt the mesh density
to the local feature size of the (approximate) intersection curves. 
This would significantly improve results, particularly at higher levels of refinement.

\subsection{Reprojection Constraint}

\begin{figure}[htb]
 \centering
  \includegraphics[width=\linewidth]{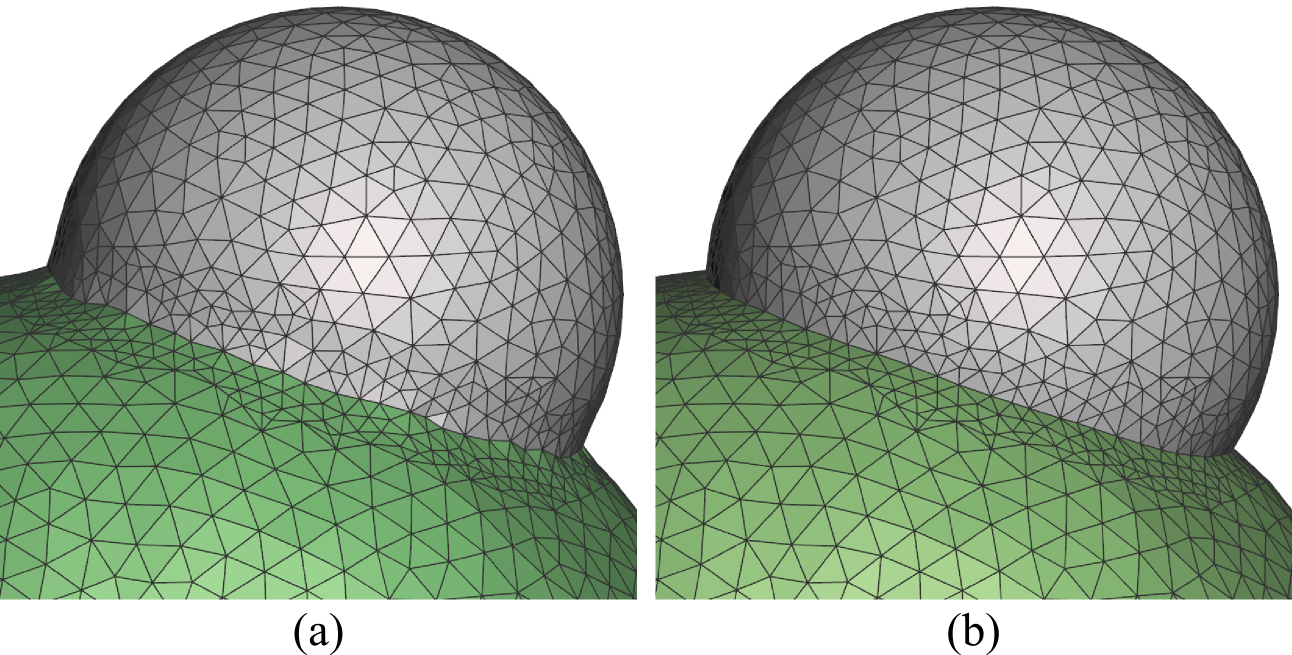}
  \caption{\label{figApproxPrecise} Our basic method produces (a) an approximate Boolean.
By adding a reprojection constraint, we can (b) approximate the intersection curve up
to a user-defined tolerance. }
\end{figure}

So far, our Boolean as described is only approximate, as the zippering step will pull the
boundary loops away from the input surfaces as they evolve towards each other. Instead, we would 
like to constrain this evolution so that rather than moving freely through 3D space, each loop can only
``slide'' along the original surface it came from. To accomplish this we employ 
a reprojection step, where after any loop vertex $\ev$ is moved, we immediately
project it to the nearest point on the input mesh $\ev^s$. 
Although this slows convergence somewhat, in most cases it results in
a high-quality intersection curve, see Figure~\ref{figApproxPrecise}.

\subsection{Sharp Edge Constraints}

\begin{figure}[htb]
 \centering
  \includegraphics[width=\linewidth]{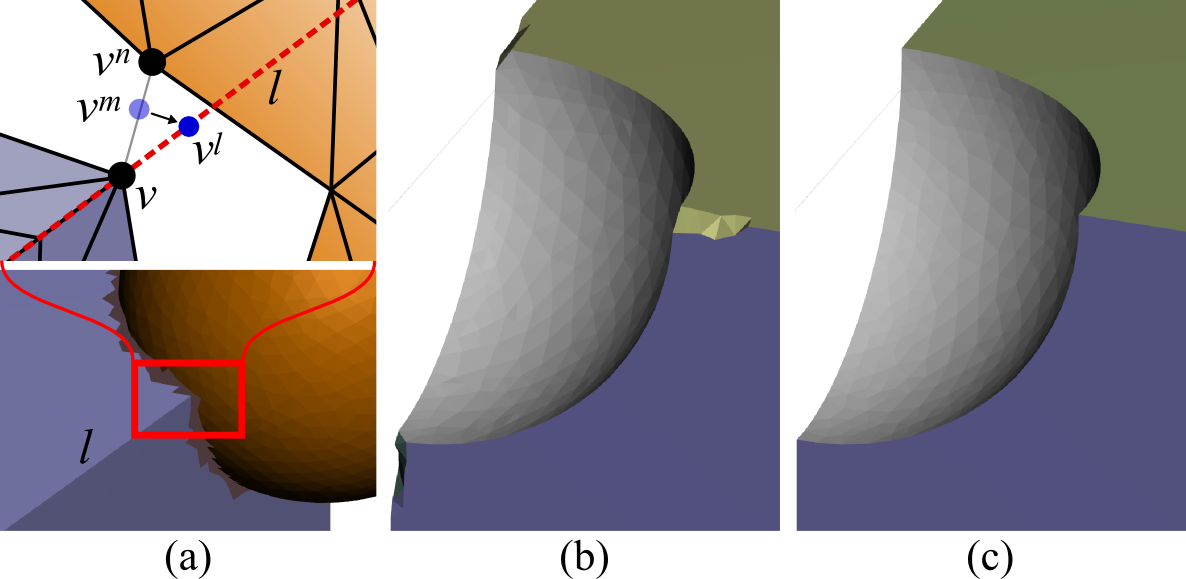}
  \caption{\label{figProjection} We (a) constrain vertices on the detected crease curve $l$
by first taking a step towards $v_n$, then reprojecting onto $l$. 
Combined with remeshing constraints, the smoothing of sharp edges in (b) is
mitigated (c) by this reprojection step. }
\end{figure}

In the discussion of mesh refinement above, we mentioned that some edges represent critical features
on the mesh surface, and we introduced constraints to prevent the loss of such features during refinement. 
Similarly, we must be careful during the reprojection step of our Boolean algorithm.
Figure~\ref{figProjection}b shows a simple case where a na\"{i}ve implementation of our Boolean will mangle sharp edges
in the neighborhood of the intersection. This is due not only to careless edge flips and smoothing (which can be remedied using
the refinement constraint mechanism introduced earlier), but also to the vertex movement during the reprojection step. To 
preserve sharp edges we must constrain this vertex movement.

Assume we have a set of constraint edges forming one or more 3D polylines.
Each vertex $\ev$ initially lying on a constraint polyline $l$ is associated with $l$.
During the projection step for $\ev$, instead of moving the vertex to the
nearest on-surface point $\ev^s$, we find the nearest point $\ev^l$ on the polyline $l$ and
move the vertex there.

\subsubsection{Corner Gaps}

\begin{figure}[htb]
 \centering
  \includegraphics[width=.7\linewidth]{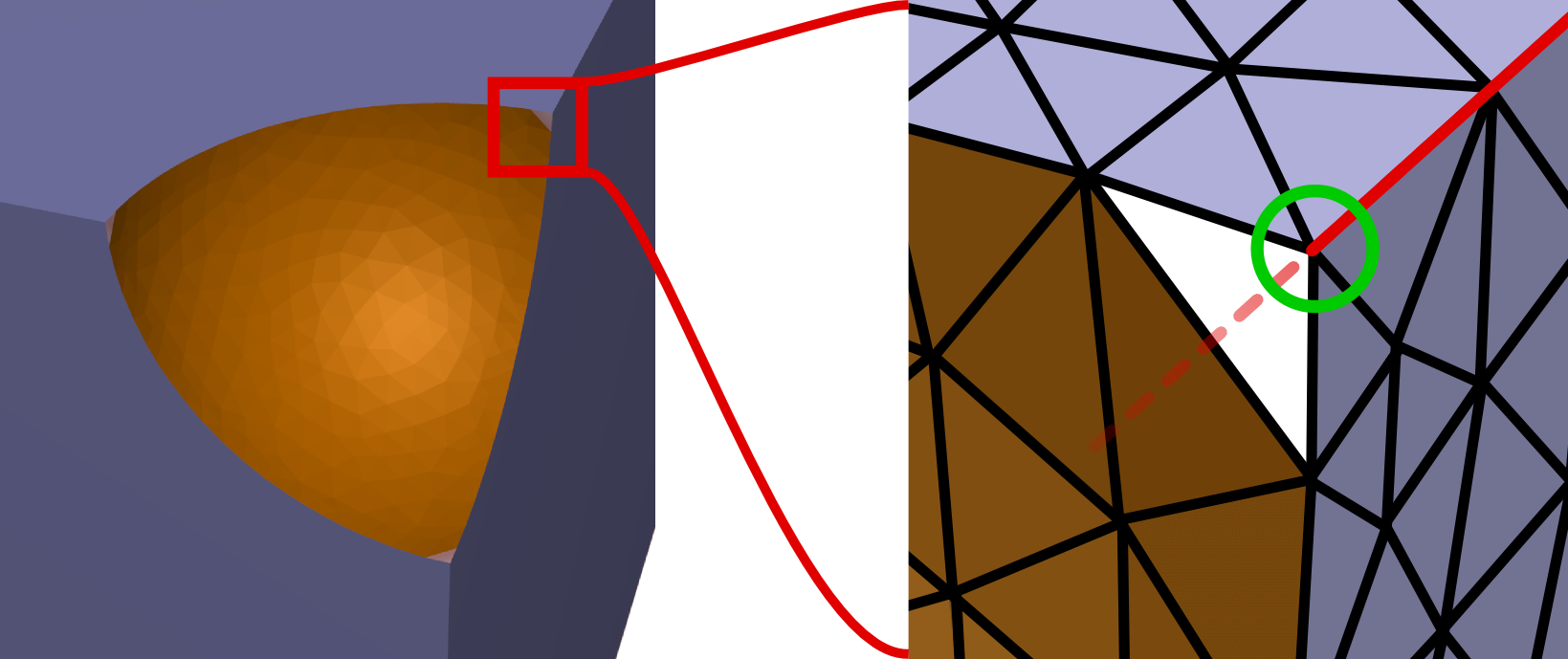}
  \caption{\label{figCornerHole} The circled vertex is constrained to the dashed red line, but the
opposing edge is not long enough to split, leaving a hole. }
\end{figure}

Although our zippering technique is highly robust, introducing sharp edge constraints
brings with it an additional complication, as illustrated in Figure~\ref{figCornerHole}.
Without the constraint, the corner vertex would move towards one of its
adjacent neighbors, leading to an edge collapse that would remove the extra vertex.
However, the edge constraint prevents this from happening, resulting in a triangular hole. 
Once the evolution has converged, this case is easily identifiable because the 
constrained vertex will not be ``nearest'' to any vertex on the opposing loop,
and so we can directly append the missing triangle.

\subsubsection{Border Edges}

A second complication of our approach to maintaining sharp edges occurs if
the step that deletes triangles in the neighborhood of the intersection
discards triangles adjacent to a sharp edge. 
When this occurs, the vertices constrained to remain on the sharp edge cannot move towards the
intersection crease, resulting in a failure to converge.

\begin{figure}[htb]
 \centering
  \includegraphics[width=\linewidth]{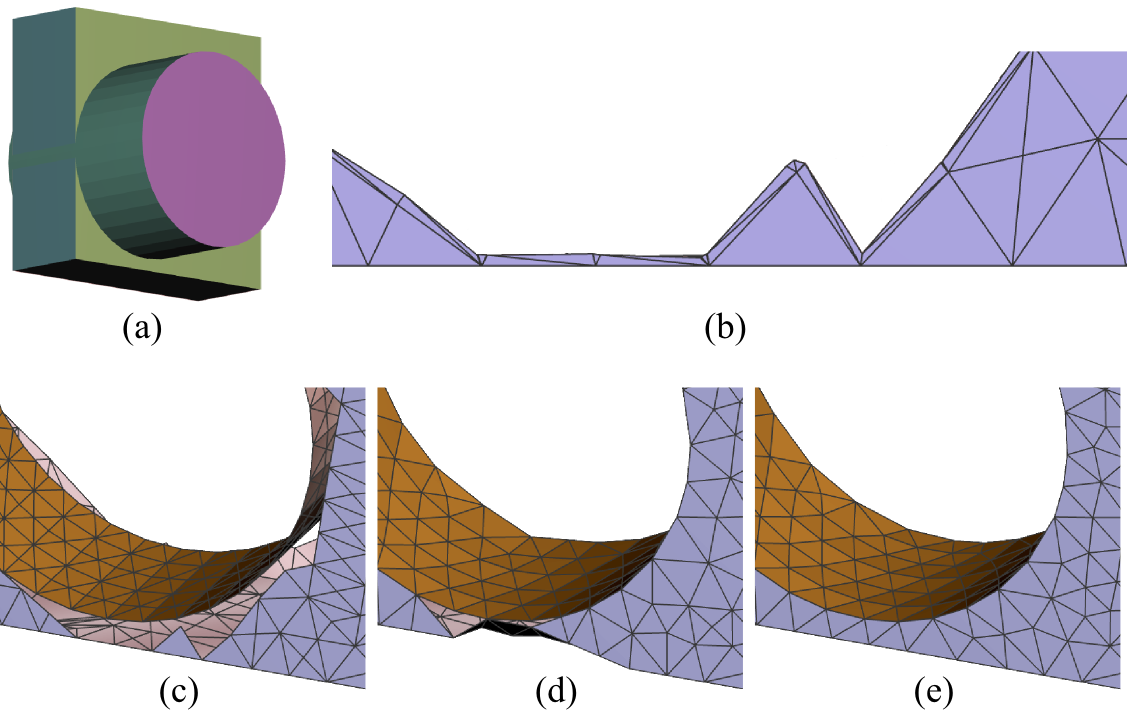}
  \caption{\label{figBorderStrip} In case (a), when the intersecting triangles (c) are
deleted, an edge of the original box lies on the open boundary. 
Preserving this edge as a sharp constraint results in a hole (d).
If we append a strip of border triangles (b), the result converges (e). }
\end{figure}

In most cases adaptive refinement will resolve this issue, as eventually the subdivision level
will be such that the faces adjacent to the crease will not be deleted.
However, we can accelerate the process by adding a small strip of triangles around the cut border. 
This gives us a set of free edges/vertices on the boundary, which can safely be evolved.
See Figure~\ref{figBorderStrip} for an example.

\subsection{Postprocessing}

\begin{figure}[htb]
 \centering
  \includegraphics[width=\linewidth]{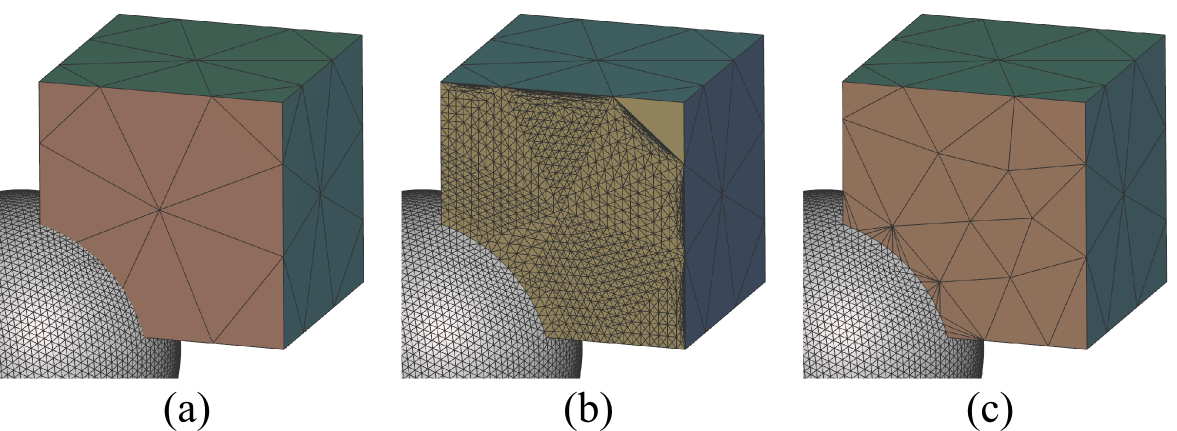}
  \caption{\label{figSimplify} When performing Booleans with (a) low-resolution meshes,
our approach will (b) significantly refine triangles in the (mesh-topological) neighborhood of
the intersection. We (c) reduce the result automatically as a post-process. }
\end{figure}

One drawback of our method described thus far is that it can drastically
increase the triangle density in regions of intersection. 
For example, a so-called ``low-poly'' mesh is not really an adaptive mesh surface,
as each triangle is critical to the overall shape. 
In such cases our Boolean will subdivide many times in the intersection neighborhoods,
which may be undesirable for computational or aesthetic reasons. 
Hence, similar to \cite{Pavic11}, we include automatic simplification of the intersection 
region as a postprocess. Figure~\ref{figSimplify} demonstrates this capability.

\section{Evaluation}

We have experimented extensively with our techniques, and they have also been in widespread
use in publicly-available free software (See section~\ref{secInteractive}). 
In this section we briefly detail some of our experiences.
We emphasize that our method is based on the assumption that the input meshes are
adaptive mesh surfaces, i.e.\ where the sampling rate is high relative to the scale of salient shape features.
If this does not hold, then a polyhedral Boolean may be more appropriate 
(although our method does often produce good results in these cases, e.g., see Figure~\ref{figSimplify}).

\subsection{Robustness}

Robustness is a major problem for mesh Booleans. The main challenge is degenerate configurations, 
such as coplanar triangles, or a vertex of one triangle lying on the face of another.
In floating-point arithmetic, such cases can be unstable: for a patch of co-planar triangles,
some may be identified as intersecting and others will not. After cutting the existing faces and
removing internal regions, the result may be non-manifold, and stitching the sets of remaining
faces may be ambiguous. 

\begin{figure}[htb]
 \centering
  \includegraphics[width=\linewidth]{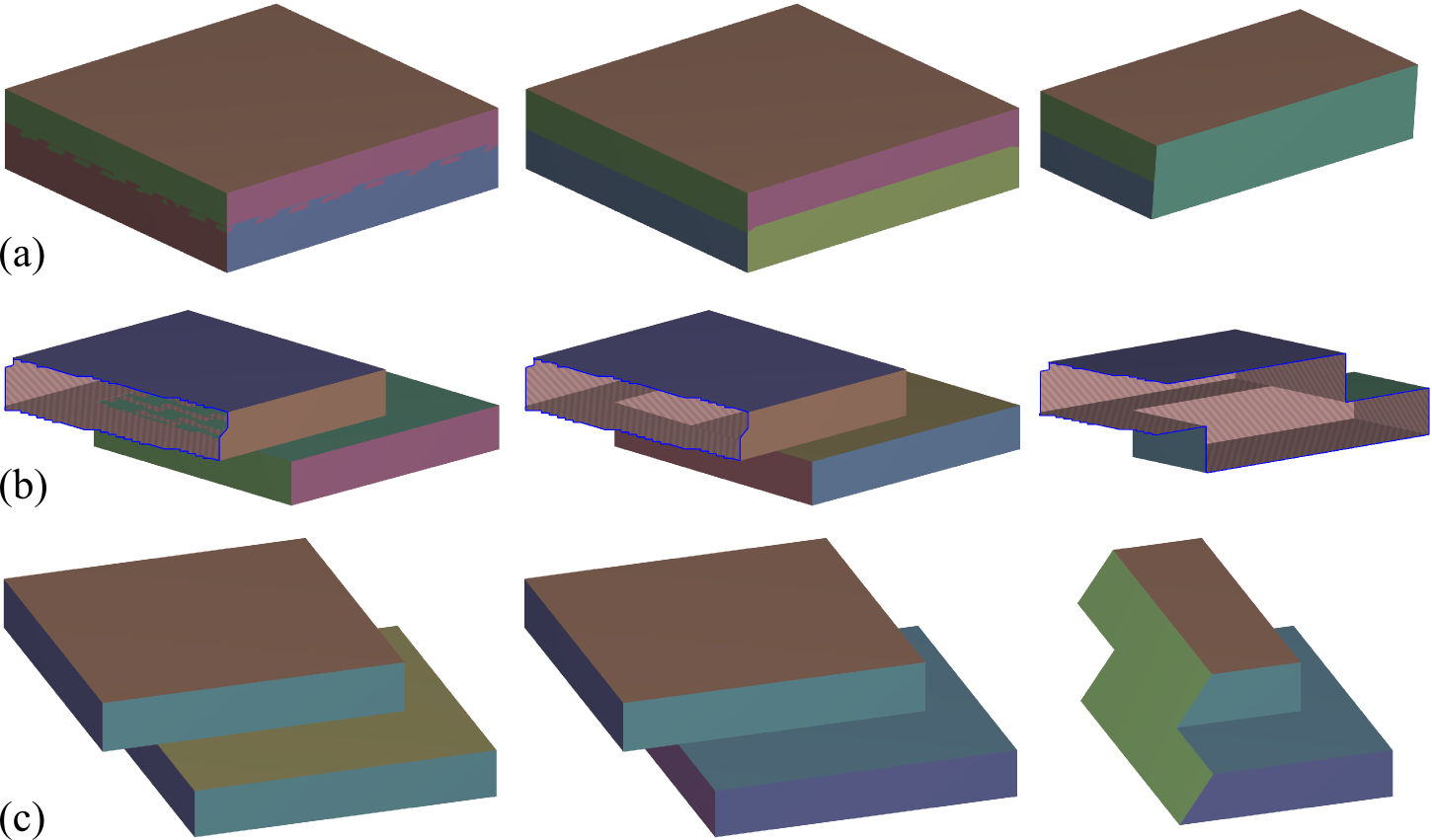}
  \caption{\label{figCaseCompare} Cases where (a) Carve, (b) CGAL, and (c) Cork fail (either crash
or produce nonmanifold output), and our method succeeds. Left column shows the two objects,
middle shows the successful result computed by our method, and Right column shows planar
cuts through the result, demonstrating that they are solid (top,bottom) or manifold (middle), as expected. }
\end{figure}

When attempting to compute exact triangle mesh intersections, as many existing Boolean libraries attempt to do, 
increasing reliability often comes at the expense of performance.
The CGAL polyhedral Booleans are generally considered the most robust,
but are extremely slow for large meshes. 
We also experimented with Carve~\cite{Carve} and Cork~\cite{Cork}, two open-source mesh 
Boolean libraries used in commercial products and by other research projects. 
For each library, we could find many cases where that method failed and ours was successful (see Figure~\ref{figCaseCompare}). 
However, we could also find cases where our current implementation failed,
and CGAL, Cork, or Carve succeeded.

\begin{figure}[htb]
 \centering
  \includegraphics[width=\linewidth]{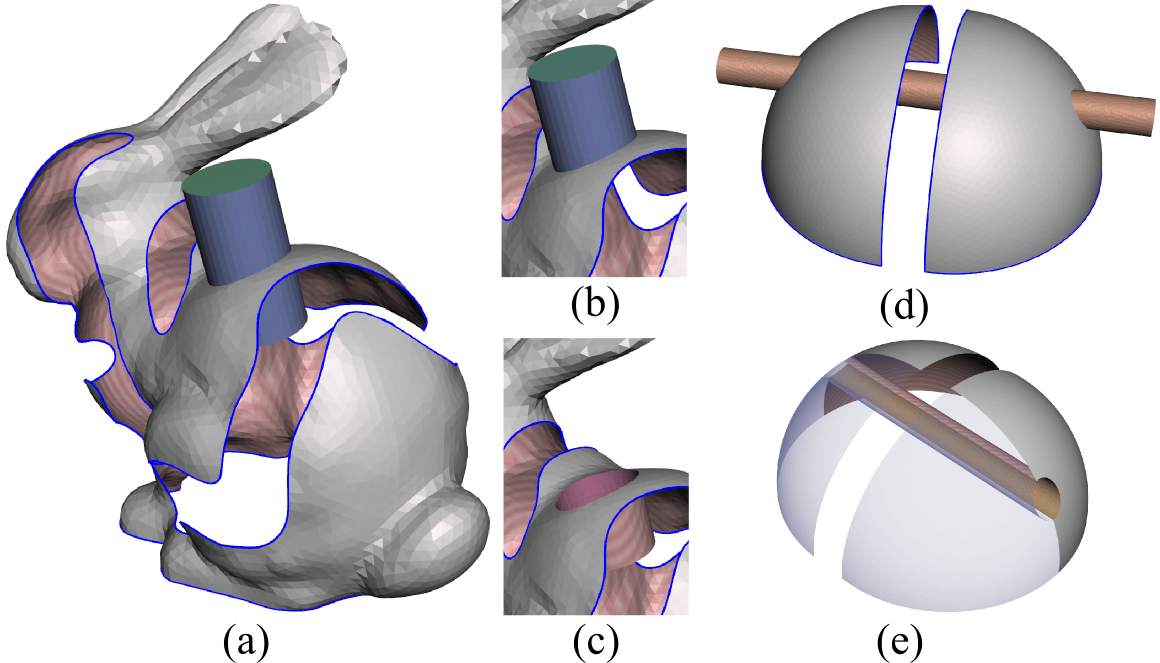}
  \caption{\label{figHoley} A cylinder intersecting with an object with complex open boundaries (a)
is (b) added and (c) subtracted. In an even more extreme case, we subtract (d) a cylinder
and show via cut-away (e) that the correct topology and orientation is produced. }
\end{figure}

Many mesh Boolean techniques assume the input meshes are closed (no boundary edges).
This is the ideal case, however in practice many input meshes will not be closed. 
One benefit of our local-remeshing-based approach is that we do not depend on mesh
properties outside the intersection regions, except for the ray-intersection tests used to
statistically determine point containment (Section~\ref{secBasicBool}).
Hence, we can perform ``Boolean'' operations with meshes that are not remotely solid (Figure~\ref{figHoley}).
Only the Cork library was capable of similar operations.

The most common failure we encountered was due to our
iterative refinement not proceeding far enough to resolve some geometric ambiguity.
As we use uniform refinement, we may require extreme triangle counts in near-coplanar cases.
Because we capped the refinement level, the operation would abort before reaching a suitable triangle density.

\begin{figure}[htb]
 \centering
  \includegraphics[width=\linewidth]{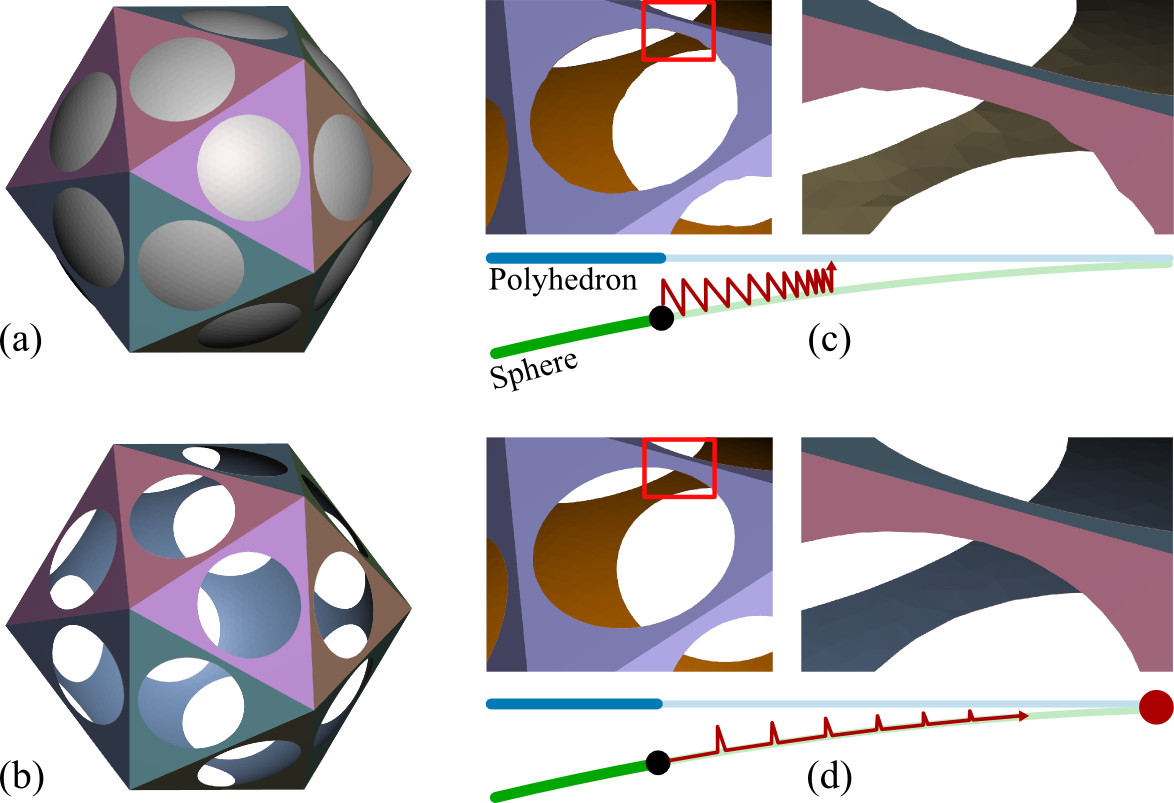}
  \caption{\label{figIntrCurves} Subtracting a sphere from a polyhedra (a,b) produces many
near-coplanar intersection regions. Our projection strategy will converge very slowly in this
case because (c) the zipper-step directions are near-perpendicular to the projection surfaces
(diagram shows cross-section perpendicular to sharp edge, red line shows path of evolving point).
By (d) also stepping towards the per-triangle intersection segments, the result  is improved. }
\end{figure}

Finally, as two input surfaces meeting at some intersection region approach
co-planarity, it takes progressively longer for the boundaries to converge.
The boundary vertices take steps that are nearly parallel to the local normals 
of their projection surfaces, and then are projected back in nearly the same direction (Figure~\ref{figIntrCurves}).
To avoid this, we can compute a set of approximate intersection line segments between triangles,
and include steps towards this shared intersection curve in our evolution.
Note that this does require computing per-triangle intersection segments, however
it does not require chaining these segments together into consistent loops.

\subsection{Coplanar and Near-Coincident Regions}

\begin{figure}[htb]
 \centering
  \includegraphics[width=\linewidth]{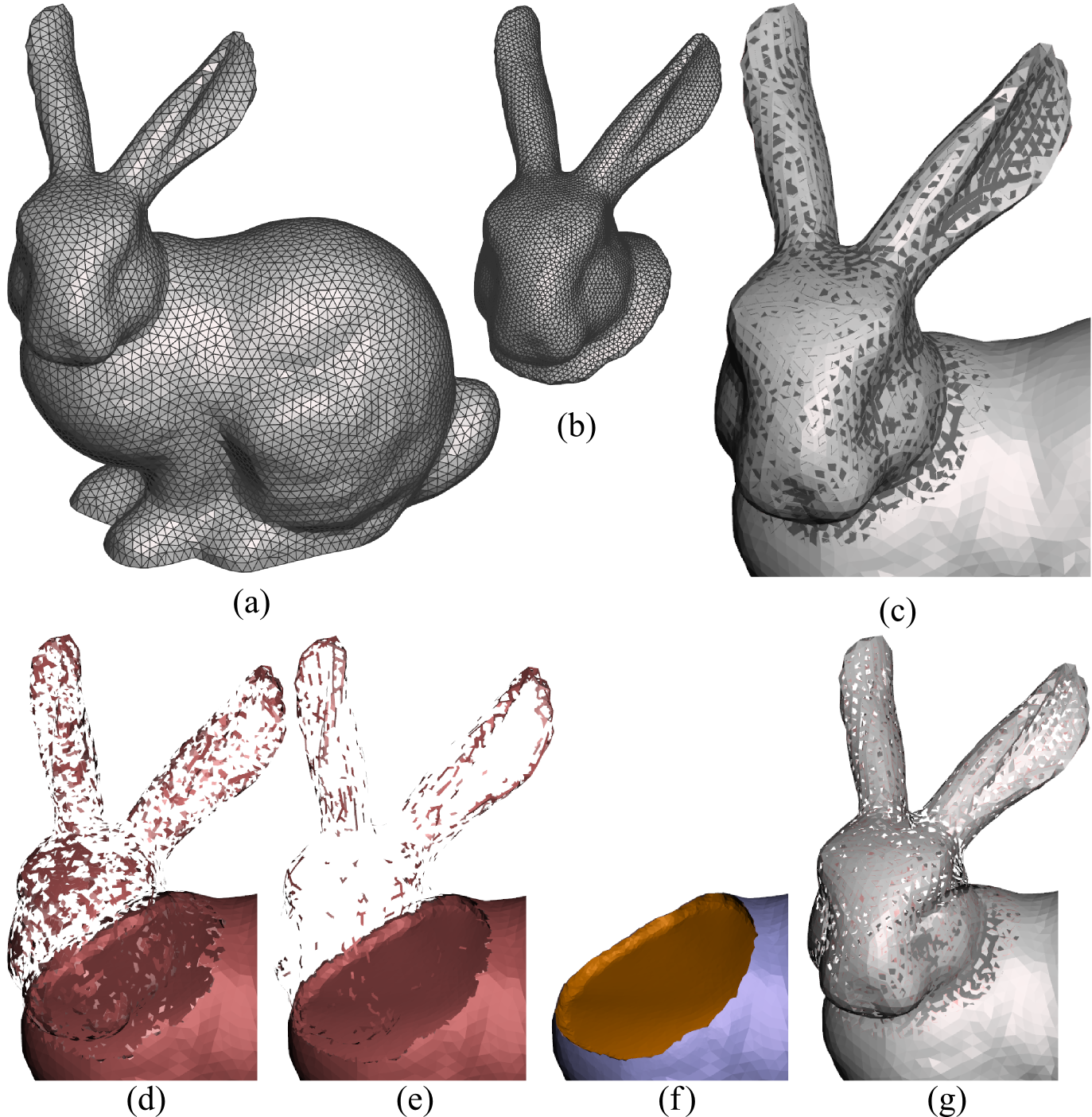}
  \caption{\label{figCoincident} Here we remove the head of the bunny (a), close the base, and remesh. 
The result (b) has a large number (c) of near-coplanar intersections with the original. 
Subtraction produces (d) many tiny components. Increasing tolerance in the intersection test produces (e) fewer
intersections, and then (f) the desired approximate Boolean result. Cork (g) produces a mesh which, though
manifold, is not particularly useful in most practical applications.}
\end{figure}

Co-planar regions, or near-coincident curved regions, are perhaps the most problematic special case
for mesh Booleans. In the curved case the mesh is often not precisely coincident, but
from the standpoint of the user, is the ``same shape'', for example two spheres with different tessellations. 
This is a case that is highly problematic for polyhedral Booleans based on precise intersection testing,
as shown in Figure~\ref{figCoincident}. 
If the operation does not fail entirely, the output is generally not what the user imagines, and often may be non-manifold. 

One significant advantage of our approach is that we have the freedom to be aggressive in identifying intersecting triangles
For example, we can incorporate a tolerance by intersecting thickened triangles.
With a sufficiently-large threshold all the noisy triangles are discarded and we 
can zipper the remaining free boundaries.
Figure~\ref{figCoincident} shows an example of this.

\subsection{Performance}
\label{secPerformance}

\begin{figure}[htb]
 \centering
  \includegraphics[width=\linewidth]{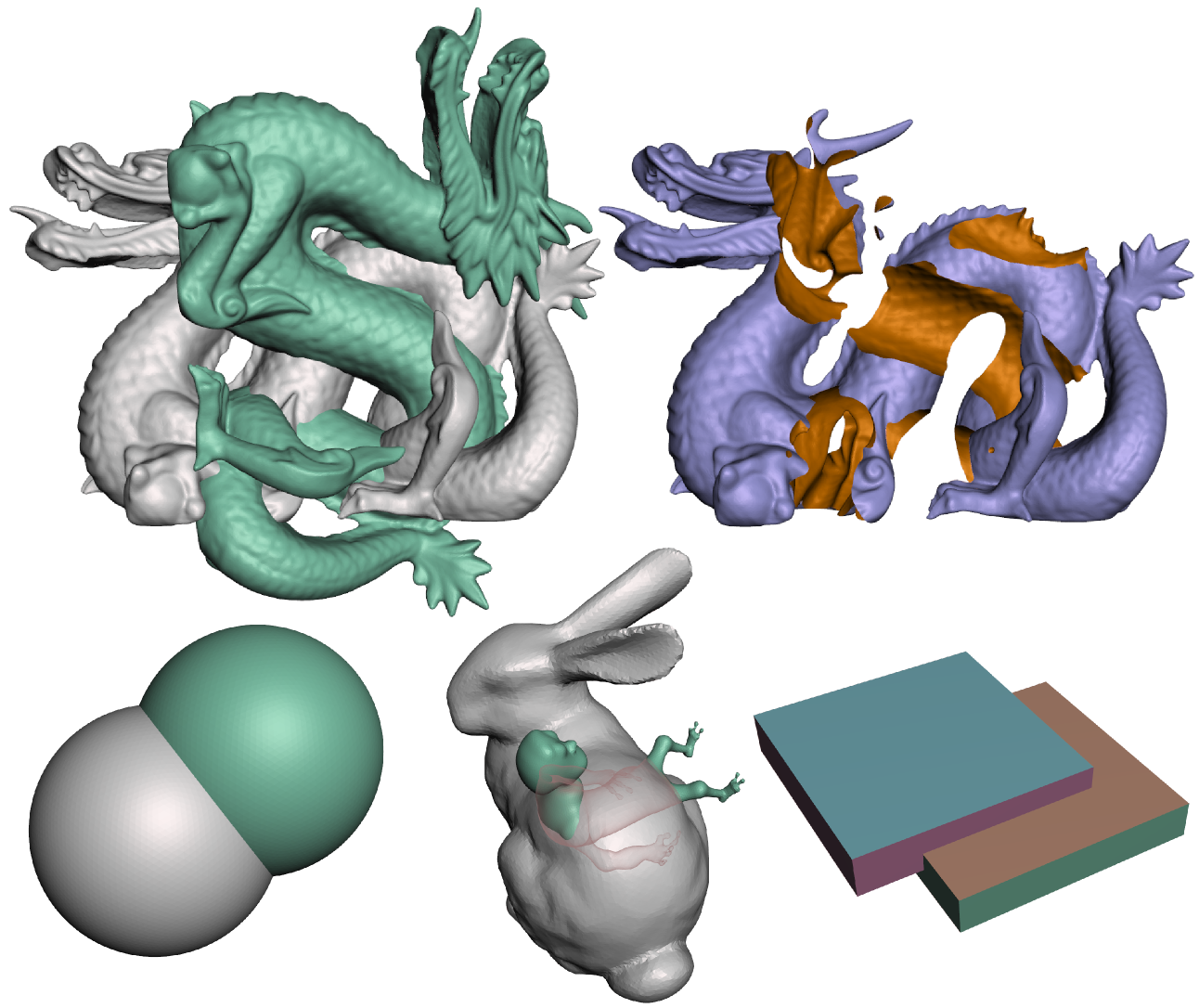}
  \caption{\label{figComplex} The test cases we used for performance profiling. The
triangle counts of each mesh are 676340 (Dragon), 12600 (Sphere),
22600 (Box), 80888 (Bunny), and 24760 (Gremlin). }
\end{figure}

As we noted, during the iterative zippering process we can 
reduce (or increase) the desired mesh density at the boundary to trade speed for precision.
Figure~\ref{figComplex} shows a zoomed-in region of the Dragon case (Figure~\ref{figComplex}),
with increasing target edge length. 
Note the highly regular edge lengths along the intersection boundary.
In our analysis, we observed standard deviations of less than $20\%$ of the
target edge length for the zippered boundary edges. 
This is in sharp contrast to many exact Boolean methods, which necessarily
produce large numbers of tiny edges and near-degenerate triangles
along the intersection curves. Although this can be repaired
with post-process remeshing, such remeshing usually done without awareness
of the inputs to the Boolean and hence will produce an inferior approximation of the intersection curve.

\begin{figure}[htb]
\centering
\includegraphics[width=\linewidth]{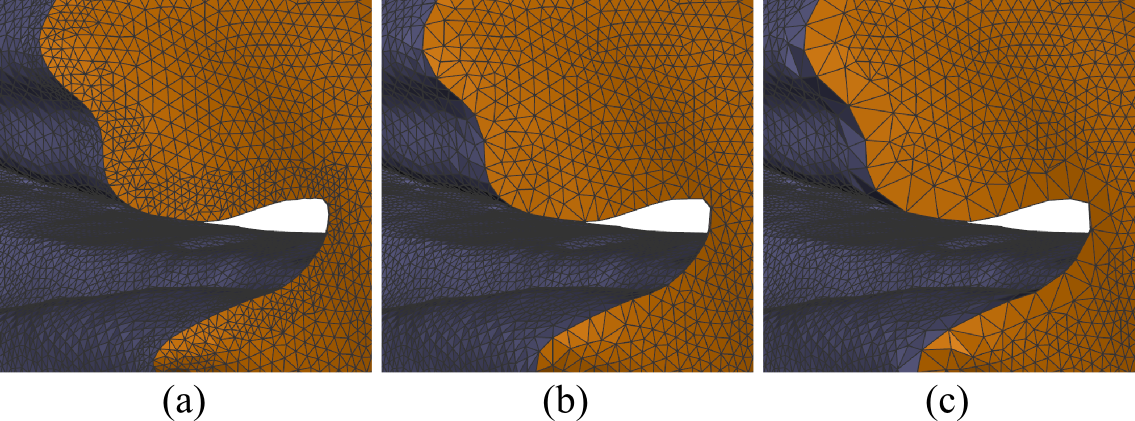}
\caption{\label{figPrecision} We can solve for the intersection curve at
variable mesh resolution, to trade accuracy for performance - default (a), 2x (b), and 3x (c). }
\end{figure}

\begin{figure}[htb]
 \centering
  \includegraphics[width=\linewidth]{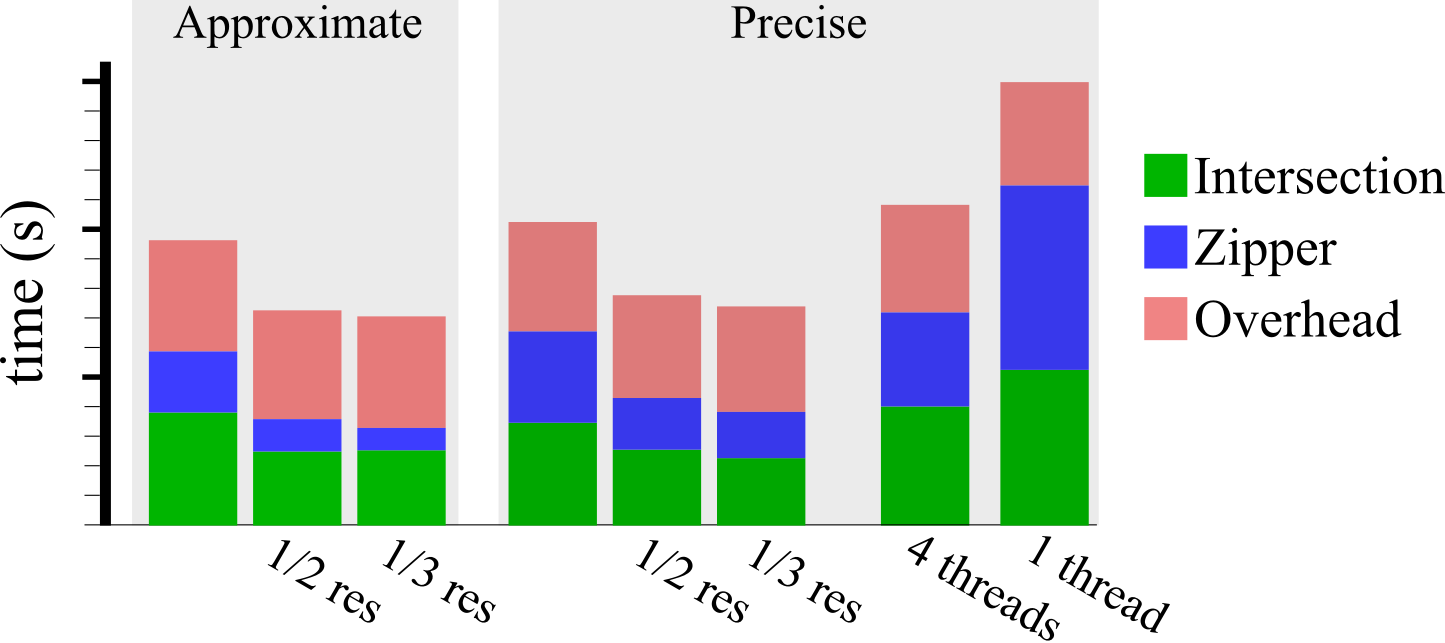}
  \caption{\label{figResTiming} Computation time for Dragon test in Figure~\ref{figComplex} on a 16-core Windows 7 64-bit workstation.
Parallel computations use 30 hyperthreads, except for rightmost bars. 
$1/2$ and $1/3$ resolution bars have reduced intersection precision, shown in Figure~\ref{figPrecision}.  }
\end{figure}

Figure~\ref{figResTiming} compares runtimes for various precision levels on the Dragon case in Figure~\ref{figComplex}. 
Our Boolean algorithm is implemented inside a general-purpose mesh modeling kernel meant for production use, and hence must support change tracking. 
The \emph{Overhead} bar includes the cost of making copies of the input meshes and journaling triangle deletions,
which are constant regardless of the solution accuracy.
The \emph{Intersection} bar includes identifying the intersection strips and determining loop correspondences, and
is also relatively constant, but does happen more quickly on lower-resolution meshes.
Finally the \emph{Zipper} bar is the time to solve the constrained zippering problem. 

The zippering process is essentially $O(N)$ in the number of boundary vertices, and hence
benefits most from reduced resolution. 
Zippering is roughly twice as fast in the Approximate solution, as the reprojection
step is quite expensive. 
The join/reproject steps are also trivially parallelizable, so the Zipper step
sees the largest gains when more CPU cores are available.

\begin{figure}[htb]
 \centering
  \includegraphics[width=0.95\linewidth]{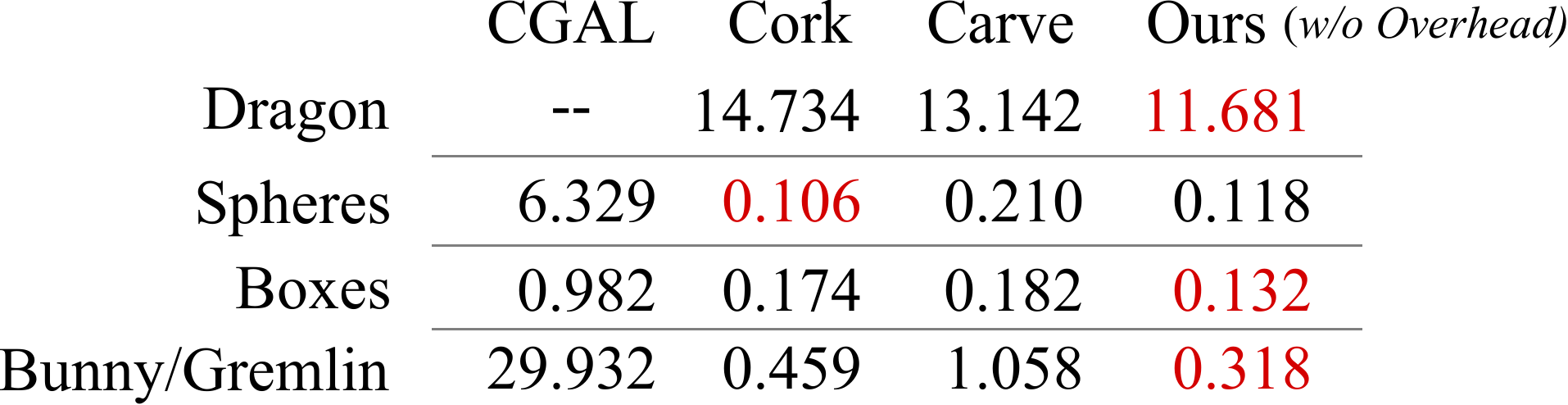}
  \caption{\label{figTiming} Computation times (s) for Boolean operations
shown in Figure~\ref{figComplex}, with CGAL, Cork, Carve, and our method.
The time to convert to/from the mesh representation required by each library
is \emph{not} included. This is not significant for Cork and Carve, but the
Nef conversion used by CGAL can take as much (or more) time that the Boolean operation. }
\end{figure}

Figure~\ref{figTiming} compares our computation times with the CGAL, Cork, and Carve
libraries, for the four examples shown in Figure~\ref{figComplex}. 
These tests were performed on a 2014 quad-core 64-bit Macbook Pro, and hence the Dragon
case takes slightly longer than in Figure~\ref{figResTiming}. 
We find that our method is generally competitive with these other libraries, although
none of these cases required multiple refinement iterations to resolve loop-matching failures.
Our computation times can grow significantly if this is necessary.

\subsection{Interactive Use}
\label{secInteractive}

\begin{figure}[htb]
 \centering
  \includegraphics[width=\linewidth]{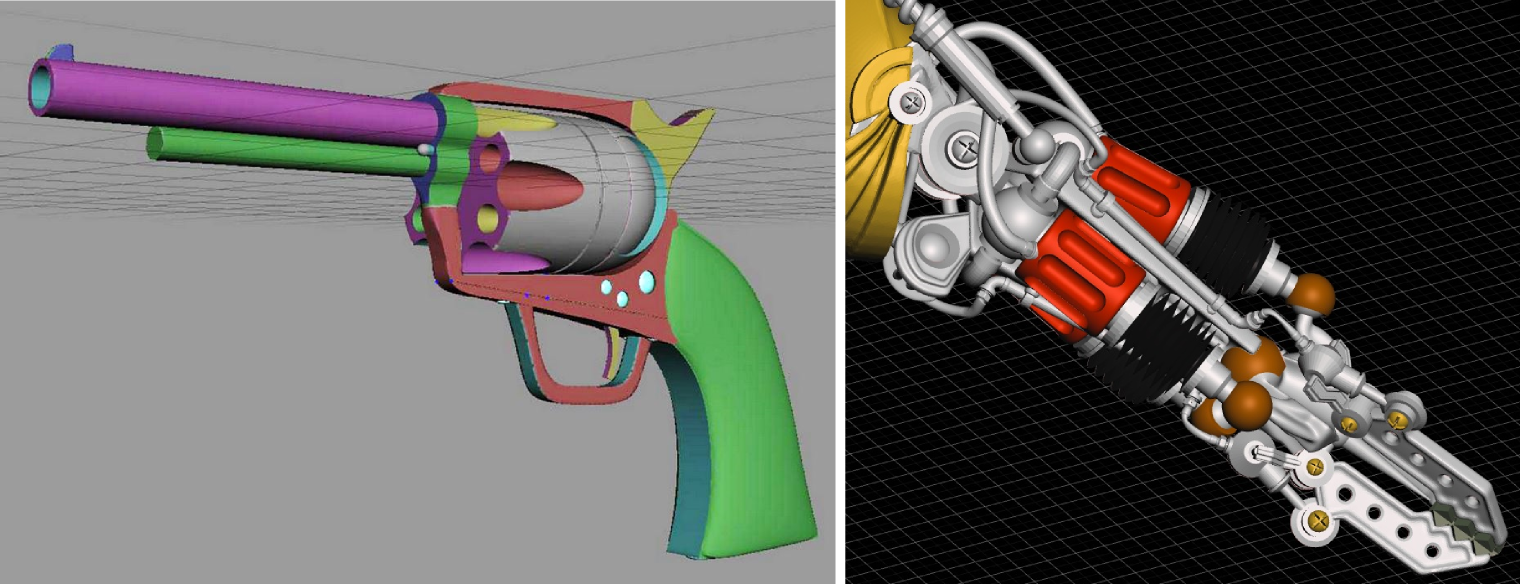}
  \caption{\label{figUsers} Models created by users that involved many Boolean operations.}
\end{figure}

A version of our adaptive mesh Booleans are implemented in Autodesk Meshmixer,
and have been in active use since 2011. As a result, we have collected extensive feedback from users,
and some examples are shown in Figure~\ref{figUsers}.

The current Meshmixer implementation does not include the sharp-edge enhancements
we described above, and also does not increase refinement if the loop-matching fails,
so users do experience frequent failures. 
However, we have observed that many users learned to troubleshoot 
Boolean failures by manually applying similar refinement strategies. 
Meshmixer includes local remeshing capabilities, and these users realized that
simply by adding more triangles in the intersection regions, they could ``fix'' a failing Boolean.
One user even reported that he \emph{preferred} our unreliable Boolean operation to those
available in other software. The reason was that although the alternatives had higher success rates,
when they failed, there was no clear way to identify and resolve the problem, while with
our Boolean the failures were easily fixed.

\section{Conclusions}

We have presented a method for performing approximate Boolean operations on general triangle meshes. 
Our approach is based on the premise that when a triangle mesh is being employed as a discretization
of an underlying smooth surface, no single triangle is important, allowing us to locally adapt and modify
the triangles as necessary. We call this an \emph{adaptive mesh surface}. 
Our zippering approach is a variation of adaptive front marching, and can also be used, for example, to join open 
meshes together or to repair holes in a general triangle mesh. 
Our constrained zippering approach to adaptive mesh Booleans produces high-quality triangulations
that can be tuned to be fast and approximate, or more precise by spending more computation time.

Performance is competitive with modern polyhedral Boolean libraries, and we expect to see 
significant performance gains in the future with a more optimized mesh refinement implementation.
In addition it may be possible to incorporate the efficient GPU techniques used by
Chentanez et al.~\cite{Chentanez15} which would further improve the capabilities of our approach.

\begin{figure}[htb]
 \centering
  \includegraphics[width=\linewidth]{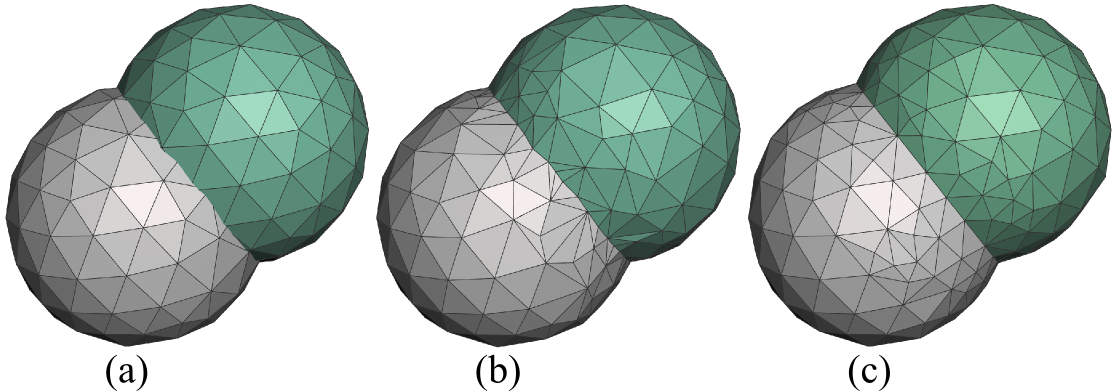}
  \caption{\label{figReproj} Two (a) coarse mesh approximations of spheres produce have (b) an
irregular intersection curve. However if the underlying analytic spheres are known, then we
can solve (c) for the precise intersection curve by constraining vertices to the analytic geometry
during our boundary evolution. }
\end{figure}

Another interesting aspect to explore is that we do not have to use the input mesh as the 
projection constraint. Consider the case where we know underlying analytic geometry for
one or both of the given meshes. We can reproject onto the analytic geometry, rather than the approximate
meshes, to produce more accurate Boolean results. Figure~\ref{figReproj} shows a simple
example with two spheres. This approach will allow us to, for example, compute a Boolean
operation between a NURBS model and an Implicit surface, without needing to actually
solve the analytic intersection problem. Although one might argue that the result will
only be an approximation, meshes remain the primary representation for
both rendering and manufacturing, and so high-quality meshes that are accurate 
up to a user-defined tolerance are in fact all that is needed in practice.

\bibliographystyle{model3-num-names}
\bibliography{BooleansSMI2016}







\end{document}